\documentclass[reprint,amsmath,amssymb,aip,apl,superscriptaddress,showkeys]{revtex4-2}
\usepackage[T1]{fontenc} 

\usepackage{graphicx}               
\usepackage[colorlinks=true,linkcolor=blue,citecolor=black,urlcolor=black]{hyperref} 

\begin{document}

\title{Chirality-Induced Spin Splitting in 1D InSeI}

\author{Shu Zhao}
\affiliation{School of Materials Science and Engineering, Zhejiang University, Hangzhou 310027, China}
\affiliation{Key Laboratory of 3D Micro/Nano Fabrication and
  Characterization of Zhejiang Province, School of Engineering,
  Westlake University, Hangzhou 310030, China}
\affiliation{Institute of Advanced Technology, Westlake Institute for Advanced Study, Hangzhou 310024, China}

\author{Jiaming Hu}
\affiliation{Key Laboratory of 3D Micro/Nano
  Fabrication and Characterization of Zhejiang Province, School of
  Engineering, Westlake University, Hangzhou 310030, China}
\affiliation{Institute of Advanced Technology, Westlake Institute for Advanced Study, Hangzhou 310024, China}

\author{Ziye Zhu}
\affiliation{Key Laboratory of 3D Micro/Nano
  Fabrication and Characterization of Zhejiang Province, School of
  Engineering, Westlake University, Hangzhou 310030, China}
\affiliation{Institute of Advanced Technology, Westlake Institute
  for Advanced Study, Hangzhou 310024, China}
  
\author{Xiaoping Yao}
\affiliation{Key Laboratory of 3D Micro/Nano
  Fabrication and Characterization of Zhejiang Province, School of
  Engineering, Westlake University, Hangzhou 310030, China}
\affiliation{Institute of Advanced Technology, Westlake Institute
  for Advanced Study, Hangzhou 310024, China}

\author{Wenbin Li}
\email{liwenbin@westlake.edu.cn}
\affiliation{Key Laboratory of 3D Micro/Nano Fabrication and
  Characterization of Zhejiang Province, School of Engineering,
  Westlake University, Hangzhou 310030, China}
\affiliation{Institute of Advanced Technology, Westlake Institute
  for Advanced Study, Hangzhou 310024, China}

\date{\today}

\begin{abstract}
   Spin-orbit coupling in chiral materials can induce
   chirality-dependent spin splitting, enabling electrical
   manipulation of spin polarization. Here, we use first-principles
   calculations to investigate the electronic states of chiral
   one-dimensional (1D) semiconductor InSeI, which has two
   enantiomorphic configurations with left- and right-handedness. We
   find that opposite spin states exist in the left- and right-handed
   1D InSeI with significant spin splitting and
   spin-momentum collinear locking. Although the spin states at the conduction
   band minimum (CBM) and valence band maximum (VBM) of 1D InSeI are
   both nearly degenerate, a direct-to-indirect bandgap transition
   occurs when a moderate tensile strain ($\sim$4\%) is applied along
   the 1D chain direction, leading to a sizable spin splitting
   ($\sim$0.11~eV) at the CBM. These findings indicate that 1D InSeI
   is a promising material for chiral spintronics.
\end{abstract}

\maketitle

Chirality extensively exists in nature, originating from a symmetry
breaking where an object cannot superimpose itself through reflection
or inversion operations~\cite{Flack2003, Fecher2022, Long2020,
  Yang2021}. Chiral materials exhibit a wide range of fascinating
properties, including magneto-chiral dichroism~\cite{Train2008},
non-reciprocal transport~\cite{Tokura2018}, and novel topological
quantum phenomena~\cite{Chang2018}. In particular, chiral materials
could be promising candidates for spintronic devices, since their
chirality-dependent spin splitting allows the electrical manipulation
of spin polarization~\cite{Yang2021, Rikken2001, Yoda2015, Inui2020,
  Shiota2021, Calavalle2022}.  However, the development of
chirality-based spintronics is currently dominated by organic and
hybrid organic-inorganic materials~\cite{Naaman2019, Yang2021}.
Despite the discovery of some intriguing phenomena in these
chiral materials such as chiral-induced spin
selectivity~\cite{Naaman2019}, the lack of inorganic materials with
chirality~\cite{Inui2020, Shiota2021, Calavalle2022} still hinders the
utilization of chiral spintronic devices.

\begin{figure*}[t!]
  \centering \includegraphics[width=0.8\textwidth]{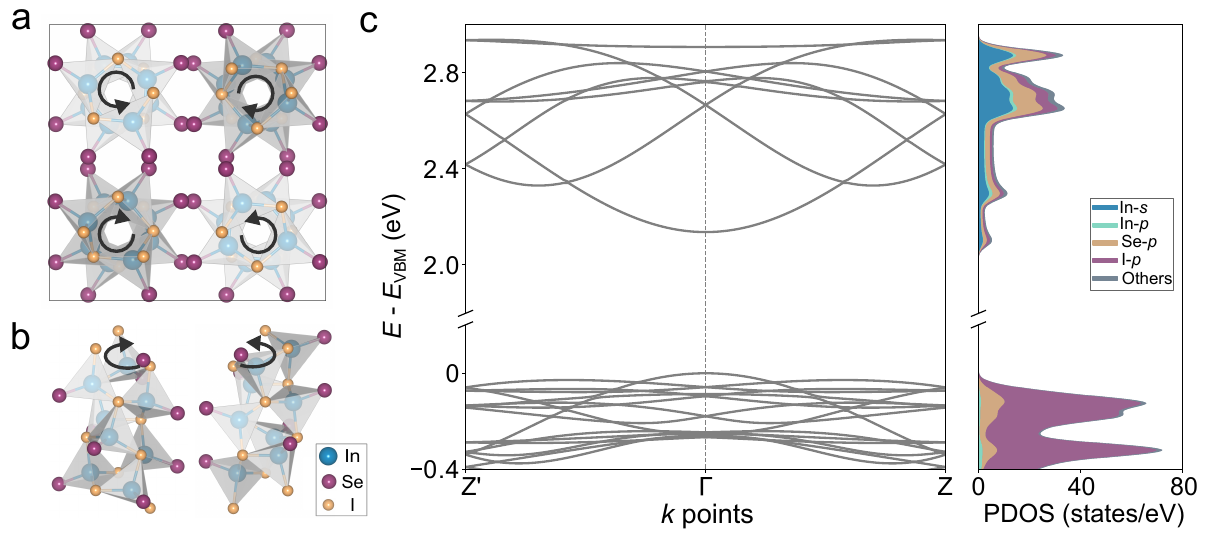}
  \caption{Crystal and electronic structures of InSeI. (a)
    The crystal structure of bulk InSeI contains two types
    (left-handed and right-handed) of chiral chains.  (b) Crystal
    structures of left- and right-handed 1D InSeI.  Blue, yellow, and
    purple balls represent In, Se, and I atoms, respectively. (c)
    Electronic band structure and projected density of states (PDOS)
    of the 1D InSeI without the inclusion of spin-orbit coupling
    (SOC), as calculated by density functional theory (DFT).}
  \label{fig:fig1}
\end{figure*}

Bulk indium selenoiodide (InSeI) has a quasi-one-dimensional structure
consisting of both left- and right-handed chiral nanochains, as
illustrated in Figure~\ref{fig:fig1}a. The synthesis of bulk InSeI and
its crystal structure analysis was first reported by Sawitzki et
al. in 1980~\cite{Sawitzki1980}. Recently, one-dimensional (1D)
nanochains of InSeI were separated from their bulk counterpart by
micromechanical exfoliation~\citep{Choi2022}, further enabling the
studies of 1D InSeI. In addition to these experimental advances,
theoretical calculations reveal that 1D InSeI exhibits a large direct
bandgap ($\sim$3.15~eV) and a moderate electron effective mass
($\sim$0.49~$m_{0}$)~\cite{Jiang2020}, indicating the potential of 1D
InSeI for nanoelectronic and optoelectronic applications. However, the
effect of chirality on the electronic structure of 1D InSeI is still
underappreciated.

In this work, through density functional theory (DFT) calculations, we
investigate the electronic and spin states of chiral 1D InSeI with
left- and right-handedness. Importantly, the band structure of 1D
InSeI exhibits significant spin splitting due to the inversion
symmetry breaking of the chiral structure. Although the spin states
(containing spin up and spin down) are nearly degenerate at the
conduction band minimum (CBM) and valence band minimum (VBM), a
sizable spin splitting ($\sim$0.11 eV) could be present at the CBM by
applying a moderate amount ($\sim$4\%) of tensile strain. The large
spin splitting at the CBM suggests that spin-polarized current with
chirality dependence could be electrically generated in the strained
1D InSeI. Therefore, our work not only elucidates the effect of
chirality on the electronic structure of 1D InSeI, but also
demonstrates the potential of 1D InSeI for chiral spintronics.

The calculations of the electronic states in 1D InSeI are performed by
DFT as implemented in the VASP code~\cite{Kresse1996}, using the
projector augmented-wave (PAW) method~\cite{Bloechl1994, Kresse1999}
and the Perdew-Burke-Enzerhof (PBE) exchange-correlation
functional~\cite{Perdew1996}. Setting the $z$ axis along the chain
direction, a vacuum thickness of 15 \AA\ across $xy$-plane is included
to avoid the spurious interaction between the helical chain and its
periodic images. A plane-wave cutoff energy of 500 eV and 1 $\times$ 1
$\times$ 12 grids of $\mathbf{k}$-points are employed, which are
sufficient to converge the total energy within 3~meV per atom.  The
unit-cell dimension along the $z$-axis and the atomic positions are
fully relaxed until the energy is converged to within 10$^{-6}$ eV and
the maximum forces on each atom are less than 0.001 eV/\AA.

The crystal structure of 1D InSeI belongs to the P$4_{1}$ (No.~76)
space group and exhibits two helical enantiomorphic configurations
with left- and right-handedness, as shown in
Figure~\ref{fig:fig1}b. The corner-sharing I-In-Se tetrahedra are
connected to form a helical structure along the $c$ axis. Each Se atom
is shared by three tetrahedra and within each tetrahedra, one In atom
links with one I atom and three Se atoms. The In-Se bond length as
obtained from DFT calculations is measured to be in the range of 2.65
to 2.7~\AA, while the In-I bond length is 2.7~\AA\ (see
\textbf{Figure~S1}). Each unit cell of the 1D InSeI nanochain contains
24 atoms (8 formula units), and the fully-relaxed unit-cell dimension
along the $z$-axis is 10.4~\AA.

We start the investigation of the electronic states of 1D InSeI by
carrying out band structure calculations without the inclusion of
spin-orbit coupling (SOC). As illustrated in Figure~\ref{fig:fig1}c,
the conduction band minimum (CBM) and valence band maximum (VBM) of 1D
InSeI are both located at the $\Gamma$ point with a DFT-PBE bandgap of
2.14 eV. Using the more accurate HSE06 functional~\cite{Heyd2003}, the
calculated bandgap of 1D InSeI is 3.06 eV (see \textbf{Figure~S2}),
which is larger than those of most other reported 1D semiconductor
materials~\cite{Peng2018, Pfister2016, Zhao2023}. From the right panel
of Figure~\ref{fig:fig1}c, it can be seen that the conduction bands
(CBs) of 1D InSeI has a significant contribution from the In-$s$ and
In-$p$ orbitals, whereas the valence bands (VBs) are dominated by the
contributions from the Se-$p$ and I-$p$ orbitals.

\begin{figure*}[t!]
  \centering \includegraphics[width=1.0\textwidth]{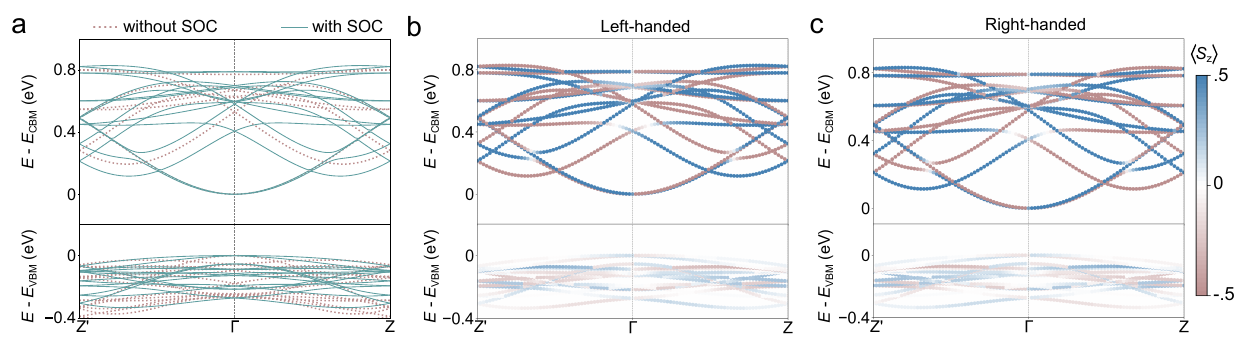}
  \caption{Chirality-dependent spin splitting of 1D InSeI. (a)
    Comparison of the calculated electronic band structures of a InSeI
    nanochain with (green lines) and without (red dotted lines) the
    inclusion of SOC. (b,c) Calculated band structures of left-handed
    (b) and right-handed (c) InSeI nanochains with the expectation
    values of the spin operator $S_{z}$ on the spinor wave
    functions. The unit of the expectation value $\langle S_{z}
    \rangle$ is $\hbar$.}
  \label{fig:fig2}
\end{figure*}

Given that the chiral 1D InSeI nanochains do not possess inversion
symmetry, SOC could have a significant influence on their electronic
structures. Hence, we next calculate the electronic states of 1D InSeI
with the SOC effect included, as well as investigating their chirality
dependence. The comparison of the band structures of 1D InSeI with and
without SOC is shown in Figure~\ref{fig:fig2}a, with the zoom-in views in the
vicinity of the valence and conduction band edges shown in
\textbf{Figure~S3}. The results indicate that 1D InSeI has a similar
direct bandgap at the $\Gamma$ point regardless of the inclusion of
the SOC effect (1.96~eV with SOC and 2.14~eV without
SOC).  Although the effect of SOC on the bandgap is insignificant, it
leads to evident band splittings in both the CBs and the
VBs. Intriguing, certain CBs exhibit much stronger
spin splitting than the others. This is because SOC requires non-zero
orbital angular momentum, and the CB states with a larger In-$p$ orbital
contribution have a stronger spin splitting than those with
the In-$s$ orbital contribution, as indicated by the k-resolved
projected density of states in \textbf{Figure~S4}.

With the inclusion of SOC effect, the 1D InSeI nanochains remain
non-magnetic and the calculated net atomic magnetic moments of In, Se,
and I atoms are all zero. However, the individual electronic states
can exhibit non-zero spin expectation values due to the spin splitting
caused by SOC. We thus calculate the expectation values of the
spin operators ($S_{x}$, $S_{y}$, and $S_{z}$) on the electronic
states of left- and right-handed 1D InSeI. The results in
\textbf{Figure~S5} show that only $\left \langle S_{z} \right \rangle$
has non-zero contribution to the spin polarization of the electronic
states, indicating that polarized spins are aligned along the chain
direction ($\pm k_z$ direction) in 1D InSeI. Such spin-momentum
collinear locking is a hallmark of the spin polarization in chiral
materials with helical structures~\cite{Calavalle2022, Yang2022}.

The band structures with the $\left \langle S_{z} \right \rangle$
values of left- and right-handed 1D InSeI are plotted in
Figure~\ref{fig:fig2}b,c. We can see that left- and right-handed 1D
InSeI exhibit opposite spin characteristics at the same wavevector and
band index, demonstrating chirality-dependent spin splitting. However,
the VBM and CBM of 1D InSeI are located very close to the $\Gamma$
point (see \textbf{Figure~S6}), where spin degeneracy is enforced by
Kramers' theorem~\cite{Melvin1974}. The rather small spin splitting at
the CBM and VBM could introduce difficulty for the generation and
detection of spin-polarized current in spintronic devices. In
addition, the calculated energy difference between the vacuum level
and the VBM is rather large (6.66 eV), exceeding the largest metallic
work function ($\sim$5.65 eV for platinum~\cite{Michaelson1977}). The
mismatch in work function would make it difficult to form Ohmic
contact with hole-conducting InSeI in practical devices. In fact,
according to the ``doping limit rule'' of wide bandgap semiconductors,
the deep VBM level of 1D InSeI indicates that its $p$-type doping
would be difficult to realize~\cite{Zhang1998}. In comparison, the
calculated energy difference between the vacuum level and the CBM has
a much smaller value of 4.52~eV. Therefore, in the following
discussion, we focus on how to induce a spin-splitting at the CBM much
larger than the thermal energy ($\sim$26 meV) at room temperature,
which would be needed for room-temperature spintronic applications.

\begin{figure*}[t!]
  \centering \includegraphics[width=1.0\textwidth]{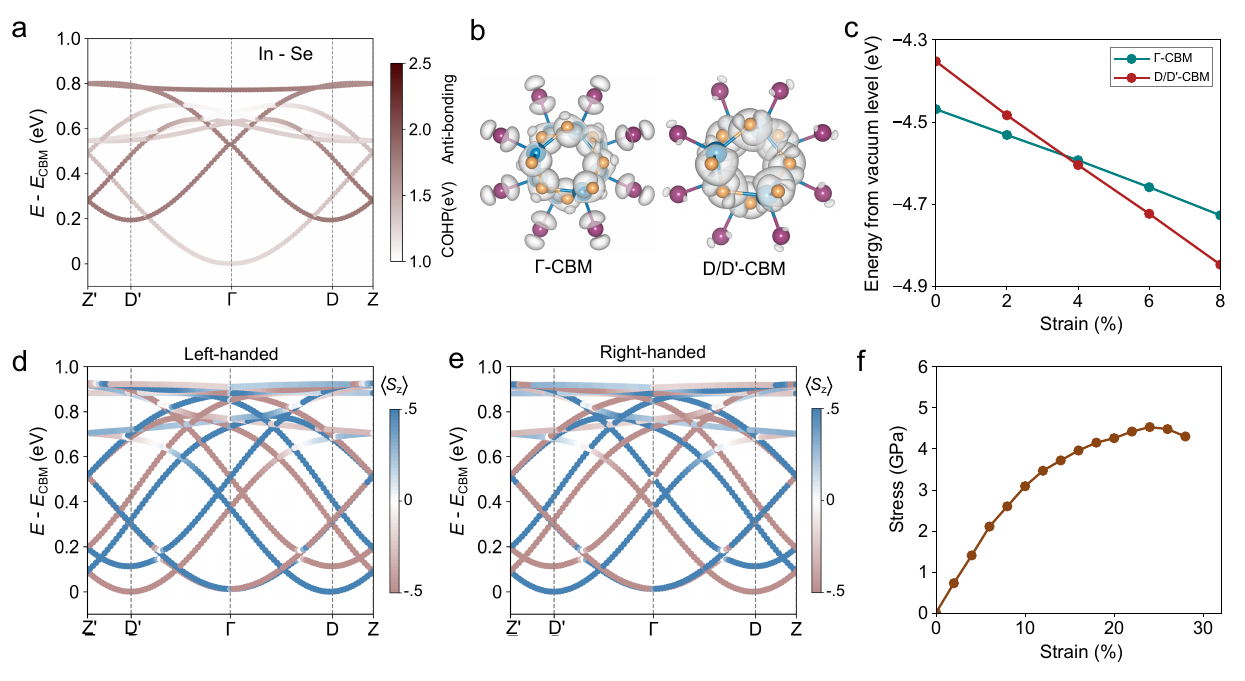}
  \caption{Strain-modulated electronic states in 1D InSeI. (a)
    State-resolved crystal orbital Hamilton population (COHP) of the
    In-Se bonds for the conduction-band electronic states. The band
    structure without the SOC effect is used for the analysis. (b)
    Partial charge densities of the conduction band minimum (CBM) at
    the high-symmetry points $\Gamma$ and $D$/$D^{\prime}$. (c)
    Strain-dependent CBM levels at $\Gamma$ and $D$/$D^{\prime}$ with
    the inclusion of SOC. (d,e) Calculated band structures with the
    $\left \langle S_{z} \right \rangle$ values on the electronic
    states of left- (d) and right-handed (e) 1D InSeI under a tensile
    strain of 4\%. (f) DFT-calculated intrinsic stress-strain curve of
    1D InSeI.}
  \label{fig:fig3}
\end{figure*}

Strain engineering has played an important role in controlling the
electronic properties of nanostructured materials~\cite{Zhu2010,
  Li2014a, Li2022}, such as the significant strain modulation of the
electronic bandgaps of 2D and 1D chalcogenides~\cite{Feng2012,
  Johari2012, Li2014b}. Indeed, tensile strain can increase the
distance between adjacent atomic sites in a crystal, resulting in
smaller electron hopping energies and thus smaller electron
bonding-antibonding splittings. This causes the energy level of an
anti-bonding orbital to move downward and that of a bonding orbital to
move upward, leading to changes in the electronic properties of
materials. From the crystal structure of 1D InSeI (see
Figure~\ref{fig:fig1}a,b), it can be seen that the In atoms and Se
atoms bond together to form the skeleton of the helical chain, while
the In-I bonds are almost perpendicular to the helical skeleton.
Therefore, the energy levels of the CBs with a stronger In-Se bonding
or antibonding character is expected to be more susceptible to tensile
strain applied along the chain direction.

Figure~\ref{fig:fig3}a depicts the wavevector-dependent crystal
orbital Hamilton population (COHP) of the In-Se bond for the
electronic states in the CBs of 1D InSeI, calculated using the LOBSTER
code~\cite{Maintz2016, Haeussermann2001}. It can be seen that the
band-energy local minimum between $\Gamma$ and $Z$/$Z^{\prime}$,
denoted by $D$/$D^{\prime}$, has a stronger In-Se anti-bonding
interaction than that of the CBM at the $\Gamma$ point. In addition,
the partial charge densities of the CBM at $\Gamma$ and
$D$/$D^{\prime}$ are calculated to visualize their wavefucntions. We
can see that, for the local band minima at $D$/$D^{\prime}$, almost
all the charge densities are distributed around the In and Se
atoms. For the CBM at $\Gamma$ point, however, significant charges
surround the I atoms. Taken the COHP and wavefunction information
together, when a tensile strain is applied along the chain direction,
the energy of the band minimum at $D$/$D^{\prime}$ is expected to be
more sensitive to the tensile strain than that at the $\Gamma$. This
will contribute to a steeper downward shift of CBM at the
$D$/$D^{\prime}$ than that at the $\Gamma$ point.

To further quantify the band energy changes, the DFT-calculated CBM
levels of a InSeI nanochain as a function of tensile strain (with the
inclusion of SOC) is shown in Figure~\ref{fig:fig3}c. Under the
applied tensile strains, which vary from 0 to 8$\%$ at a step of
2$\%$, the energy of the CBM at $D$/$D^{\prime}$ has a faster rate of
decrease than that at $\Gamma$. Once the tensile strain imposed on 1D
InSeI is larger than $\sim$4\%, the energy level of the CBM at
$D$/$D^{\prime}$ becomes lower than that of the CBM at $\Gamma$,
resulting in a direct-to-indirect bandgap transition (detailed
strain-dependent band structures are shown in
\textbf{Figure~S7}). When the transition occurs, a significant
chirality-dependent spin-splitting ($\sim$0.11~eV) emerges at the CBM,
as illustrated in Figure~\ref{fig:fig3}(d,e). The spin polarization of
the bands along the high-symmetry paths $\Gamma-Z$ and
$\Gamma-Z^{\prime}$ are opposite in left- and right-handed 1D InSeI.

We have further considered the effect of electrostatic carrier doping
on the electronic structure and strain-induced band structural change
in 1D InSeI, given that in practical devices, a finite amount of
charge carriers is present in the system. The effect of electrostatic
doping is modeled using the background charge approach, with fully
relaxed atomic positions. When the electron doping concentration
($n_e$) of 1D InSeI is 0.01 electron per formula unit ($e$/f.u.),
corresponding to $n_e \approx 1.1\times10^{20}$~cm$^{-3}$, a number
approaching the maximum electron doping concentration of
\textit{n}-type silicon~\cite{Dziewior_1977}, the change in its
electronic structure is negligible, as shown in \textbf{Figure~S8a}.
In particular, an external tensile strain of $\sim$4\% can still cause
the CBM at $D$/$D^{\prime}$ to become lower in energy than the CBM at
$\Gamma$ in the electron-doped 1D InSeI, as illustrated in
\textbf{Figure~S8b}. The result is therefore essentially the same as
that of 1D InSeI without the inclusion of additional electrons. Thus,
we conclude that the perturbation of the electrostatic doping on the
strain-induced band structural evolution of 1D InSeI is insignificant.

Although spin splitting occurs at the CBM by applying a $\sim$4\%
tensile strain, it is unclear whether 1D InSeI can withstand the level
of tensile strain. Thus, we next explore the intrinsic mechanical properties of
1D InSeI. By applying tensile strains along the chain direction from
$-1\%$ to 1$\%$ with a step of 0.2\%, the Young's modulus ($Y$) of 1D
InSeI is calculated using the formula $Y = V_{0}^{-1} \partial^{2} E /
\partial \epsilon^{2}$, where $E$ is the system energy under strain
$\epsilon$. The equilibrium volume $V_{0}$ is given by $V_{0}$ = $\pi
r^{2} L$, where $L$ and $r$ denote the length along the $z$ axis
(chain direction) and the radius of 1D InSeI in the $xy$ plane as
measured from the iodine atoms, respectively.

The calculated Young's modulus of 1D InSeI is $\sim$34~GPa, which is
much smaller than the average Young's modulus of carbon nanotubes
(1.8~TPa)~\cite{Treacy1996}. The small stiffness of 1D InSeI suggests
the potential for flexible semiconductor devices. Furthermore, we
calculate the strain-stress curve of 1D InSeI, as shown in
Figure~\ref{fig:fig3}f. The stress ($\sigma$) of the 1D InSeI is
measured using the formula~\cite{Shang2020} $\sigma = \sigma_{0}
S_0/S$, where $S_{0}$ and $S$ are the cross section ($xy$ plane) areas
with and without the inclusion of vacuum layer,
respectively. $\sigma_{0}$ is the stress along the chain direction
($z$ axis) from direct DFT output.

The critical stress and strain corresponding to mechanical failure as
determined from the stress-strain curve are 4.7~GPa and 24$\%$,
respectively. The intrinsic mechanical failure strain is greater than
the strain at which the indirect-to-direct bandgap transition
occurs. Apart from the mechanical stability, we also investigate
the dynamical and thermal stability of 1D InSeI under mechanical
strain. The phonon spectrum of 1D InSeI under a tensile strain of 5\%,
calculated using the force-constant approach implemented in the
PHONOPY code~\cite{Togo2015}, is shown in \textbf{Figure~S9}. The
absence of an imaginary phonon mode indicates the dynamical stability of
1D InSeI under a tensile strain of 5\%. In addition, \textit{ab
  initio} molecular dynamics (AIMD) simulation of 1D InSe under a
strain of 5\% at 350~K is performed using a $1\times 1 \times 3$
supercell, with a total simulation length of 5~ps at a timestep of
1~fs. \textbf{Figure~S10} shows that the helical framework and the
atomic bonds of 1D InSeI are well preserved during the AIMD run,
indicating its thermal stability. Additional calculations of the COHP
and integrated COHP (ICOHP) values of 1D InSeI as a function of the
applied tensile strain are shown in \textbf{Figure~S1}. All the
calculation results confirm that 1D InSeI maintains mechanical,
dynamical, thermal, and bonding stability under a tensile strain of at
least 5\%, which is higher than the critical strain of 4\% to induce
the direct-to-indirect bandgap transition and strong spin-splitting at
the CBM. Experimentally, it has been demonstrated that silicon
nanowires with diameters of $\sim$100~nm can be repeatedly stretched
above 10\% elastic strain at room temperature~\cite{Zhang2016},
indicating that experimental realization of the strain-induced
electronic structural changes in 1D InSeI is entirely feasible.  These
properties make 1D InSeI a useful materials system for
mechano-spintronic applications, where tensile strain could modulate
the spin polarization of the electrical current passing through the
chiral 1D material.

In summary, through ﬁrst-principles calculations and analysis, we
study the effect of chirality on the electronic states of 1D InSeI. We
find that spin-orbit coupling in the chiral 1D semiconductor induces
chirality-dependent spin splitting and spin-momentum collinear locking
in the conduction and valence bands. By applying an external
tensile strain of $\sim$4\% along the 1D chain direction, a large
spin splitting of 0.11~eV emerges at the CBM along with
a direct-to-indirect bandgap transition.  These results highlight the
potential of 1D InSeI for chiral spintronic and mechano-spintronic
applications.

We gratefully acknowledge the support by NSFC under Project
No. 62004172. The work of W.L. is partially supported by Research
Center for Industries of the Future at Westlake University under Award
No. WU2022C041. The authors thank Drs. J.-Q. Wang and C. Hu
for helpful discussions and the HPC Center of Westlake
University for technical assistance. 

\section{Supplementary Material}
Calculations of the COHP and ICOHP of 1D InSeI; band structures of 1D
InSeI under different tensile strains along the chain direction; spin
expectation values of the electronic states in 1D InSeI; the effects
of electron doping on the electronic structure of 1D InSeI; phonon
spectrum and AIMD simulation of 1D InSeI under 5\% tensile strain.

\bibliography{reference}

\end{document}